\documentclass{article}

\usepackage{amssymb,amsfonts,amsmath}
\usepackage{cite,enumerate,float}
\usepackage{color}
\usepackage{tikz}
\usetikzlibrary{arrows,snakes,backgrounds}

\def\be{\begin{eqnarray}}
\def\ee{\end{eqnarray}}
\def\nn{\nonumber}

\def\p{\partial}
\def\tr{{\rm tr}\,}
\def\Tr{{\rm Tr}\,}

\definecolor{red}{rgb}{1,0,0}
\definecolor{orange}{rgb}{1,0.5,0}
\definecolor{violet}{rgb}{0.7,0,1}

\hoffset= - 1.0in         

\begin{document}

\hfill MIPT/TH-05/24

\hfill IITP/TH-05/24

\hfill ITEP/TH-06/24

\bigskip

\centerline{\Large{
On character expansion and Gaussian regularization of Itzykson-Zuber measure
}}

\bigskip

\centerline{\bf A.Morozov and A.Oreshina}

\bigskip

\centerline{MIPT, 141701, Dolgoprudny, Russia}
\centerline{NRC “Kurchatov Institute”, 123182, Moscow, Russia}
\centerline{IITP RAS, 127051, Moscow, Russia}
\centerline{ITEP, Moscow, Russia}

\bigskip

\centerline{ABSTRACT}

\bigskip

{\footnotesize
Character expansions are among the most important approaches to modern quantum field theory,
which substitute integrals by combinations of peculiar special functions from the Schur-Macdonald family.
These formulas allow various deformations, which are not transparent in integral formulation.
We analyze from this point of view the Itzykson-Zuber integral over unitary matrices
which is exactly solvable, but difficult to deform in $\beta$ and $(q,t)$ directions.
Character expansion straightforwardly resolves this problem.
However, taking averages with the so defined measure can  look  problematic,
because integrals of individual expansion terms often diverge and well defined is only the sum of them.
We explain a way to overcome this problem by Gaussian regularization,
which can have a broad range of further applications.
}

\bigskip

\bigskip

\section{Introduction}

Matrix models \cite{MaMo} can seem trivial examples of quantum field theory, with no space-time and narrow application to physics.
However, it is long understood \cite{UFN2} that instead they focus on the non-perturbative side of the story
and highlight the main properties of non-pertutbative physics, including hidden symmetry and integrability structures --
not respected or at least not explicit in the study of Feynman diagrams, where special effort is needed to reveal them \cite{MMRS}.
After a clear understanding of the eigenvalue single-matrix model \cite{UFN3}, attention is moving closer to the
multi-matrix ones, which form a much wider world with many additional properties \cite{multMaMo,WLZZ,MMMPWZ,Kerov,MMMP}.
The key step in the move from single to multi-matrix models is the unitary calculus \cite{Munit},
originating from the theory of Itzykson-Zuber (IZ) integral \cite{IZ}.

IZ integral was originally defined as an integral over $N\times N$ unitary matrices
\be
I[X,Y]:= \frac{1}{{\rm Vol}_{U(N)}}
\int_{N\times N} e^{ \tr XUYU^\dagger} [dU]
= \frac{\det_{_{a,b=1\ldots N}} e^{ x_ay_b}}{\Delta(X)\Delta(Y)}
\label{Iint}
\ee
which depends only on the eigenvalues of $X$ and $Y$.

The goal of this paper is to shed more light on a very different representation of
(\ref{Iint}):
the character expansion formula \cite{BalKa,Munit}
which contains no reference to unitary integration:
\be
I[X,Y]:=\sum_{R: \ l_R\leq N} \frac{S_R\{\delta_{k,1}\} S_R[X] S_R[Y]}{S_R[N]}
\label{Ichar}
\ee
Denominator involves ${\rm dim}_R=S_R[I]:=S_R\{p_k=N\}$,
which sometime is also denoted by $S_R[N]$ to emphasize the dependence on $N$.
This formula can be continued to arbitrary time-variables,
\be
{\cal I}_N\{p,\bar p\}:=\sum_R \frac{S_R\{\delta_{k,1}\} S_R\{p\} S_R\{\bar p\}}{S_R[N]}
\ee
still it unavoidably depends on $N$ through the denominator.
Eq.(\ref{Ichar}) is then a restriction to the Miwa locus $p_k = \tr X^k$, $\bar p_k = \tr Y^k$
with arbitrary $N\times N$ matrices $X$ and $Y$.

A truly important advantage of (\ref{Ichar}) is the ability of straightforward $\beta$- and $(q,t)$-deformations:
\be
I^{(\beta)}[X,Y]:=\sum_{R: \ l_R\leq N}   \frac{J_R\{\delta_{k,1}\} J_R[X] J_R[Y]}{||J_R||^2\cdot J_R[N]}
\label{Ibetachar}
\ee
and
\be
I^{(q,t)}[X,Y]:=\sum_{R: \ l_R\leq N}  \frac{M_R\{\delta^*_{k,1}\} M_R[X] M_R[Y]}{||M_R||^2\cdot M_R[N^{^*}]}
\label{Iqtchar}
\ee
with the usual definitions for the $q,t$-model case \cite{qtmod}:
\be
\delta^*_{k,n} := -\frac{(1-q)^k}{1-t^{-k}}, \ \ \ M_R[N^{^*}]:=M_R\left\{p_k=\frac{1-t^{-Nk}}{1-t^{-k}}\right\}
\ee
These formulas are independent of normalization of Jack and Macdonald
polynomials  $J_R\{p\}$ and $M_R\{p\}$ \cite{MD,MOP} -- and actually  we will see below that they are exact.

These formulas can be considered  as the {\it definitions} of IZ deformations.
Their possible  relevance for the theory of WLZZ matrix models \cite{WLZZ}
was discussed recently in \cite{MMMPWZ,MOP} and \cite{LWYZ}.
In the present paper we   add some flavour of naturality to these wonderful series --
what can be useful because of the foreseeable broadness of their future applications.

We use the chance to remind that matrix models do sometime possess matrix-integral representations,
but not always, and their true definition and use are far beyond \cite{UFN3} -- despite the historical name.
Likewise we prefer (\ref{Ichar}) over (\ref{Iint}) as a {\it definition} of IZ "integral".
There are many conceptual reasons for this, but it is sufficient to mention that
$\beta$- and Macdonald deformations can neither be matrix integrals
(except for the two special points $\beta = 2^{\pm 1}$),
nor possess determinant representations,
therefore this is not a practical route towards (\ref{Ibetachar}).
The practical way is to promote (\ref{Iint}) to  (\ref{Ichar}),
which {\it is}  generalizable.
The task is therefore to make  (\ref{Ichar}) looking as natural as possible
from all what we can learn about the IZ {\it integral}  (\ref{Iint}).

\section{Straightforward proof of (\ref{Ichar})}

We begin from reminding the direct relation between (\ref{Ichar}) and (\ref{Iint}).
As explained in sec.4.2 of \cite{Munit}, there is a straightforward derivation of (\ref{Ichar})
which we now reproduce.
It relies on a few simple facts:
\begin{itemize}
\item{} For any matrix $\Psi$ Cauchy formula \cite{Cauchy} implies
\be
e^{\tr \Psi} = \sum_R S_R\{\delta_{k,1}\} S_R\{\tr \Psi^k\}
\ee
{\footnotesize
For example,  if $\Psi  =  \left(\begin{array}{cc} x_1 \\ & x_2\end{array}\right)$, then
\be
e^{x_1+x_2} = 1 + \underbrace{x_1+x_2}_{S_{[1]}[\Psi]}
\underbrace{+\frac{1}{2}}_{_{S_{[2]}\{\delta_{k,1}\}}}\!\!\!\!\!\cdot
 \underbrace{\frac{(x_1^2+x_2^2) + (x_1+x_2)^2}{2}}_{S_{[2]}[\Psi]}
\underbrace{-\frac{1}{2}}_{_{S_{[1,1]}\{\delta_{k,1}\}}}\!\!\!\!\!\cdot
\underbrace{\frac{-(x_1^2+x_2^2) + (x_1+x_2)^2}{2}}_{S_{[1,1]}[\Psi]}
+ \ldots
\nn
\ee
}

\item{} If $\Psi$ is converted into representation $R$, i.e. acquires a form of ${\rm dim}_R\times {\rm dim}_R$ matrix $\Psi_R$,
then
\be
S_R\{\tr \Psi^k\} = \Tr \Psi_R
\ee
{\footnotesize
For example, if $\Psi =\Psi_{[1]} =  \left(\begin{array}{cc} x_1 \\ & x_2\end{array}\right)$, then
$\Psi_{[2]} =  \left(\begin{array}{ccc} x_1^2 \\ & x_1x_2 \\ && x_2^2 \end{array}\right)$,
while $\Psi_{[1,1]} =  \left( x_1 x_2 \right)$,
so that
\be
\Tr \Psi_{[2]} = x_1^2+x_1x_2 +x_2^2 = \frac{(x_1^2+x_2^2) + (x_1+x_2)^2}{2} =S_{[2]}\{\tr\Psi^k\}, \nn \\
\Tr \Psi_{[1,1]} =  x_1x_2  = \frac{-(x_1^2+x_2^2) + (x_1+x_2)^2}{2}=S_{[1,1]}\{\tr\Psi^k\}
\nn
\ee
and so on.
}
\item{} The unitary-matrix integral with Haar measure is
\be
\int {\cal U}_{ab} {\cal U}^\dagger_{cd} [dU] = \frac{\delta_{ad}\delta_{bc}}{{\rm dim}_R}
= \frac{\delta_{ad}\delta_{bc}}{S_R[N]}
\ee
with ${\cal U} = U_R$ in representation $R$.

\end{itemize}
\bigskip

It remains to substitute $\Psi = XUYU^\dagger$ to get (\ref{Ichar}):
\be
I\{\tr X^k,\tr Y^k\} =  \int e^{\Psi} [dU] =  \sum_R S_R\{\delta_{k,1}\} S_R[\Psi]
= \sum_R S_R\{\delta_{k,1}\} \int \Tr X_RU_RY_RU^\dagger_R  = \nn \\
= \sum_R \frac{S_R\{\delta_{k,1}\}}{S_R[N]}\Tr X_R \Tr Y_R =
\sum_R \frac{S_R\{\delta_{k,1}\}S_R\{\tr X^k\}S_R\{\tr Y^k\} }{S_R[N]}
\ee

\section{Calogero equations
\label{Calo}}

The simplest approach to $\beta$-deformation is inspired by the differential equations,
satisfied by the IZ integral.
Since the $X$-derivative of  (\ref{Iint}) obviously produces $Y$,
it is easy to guess that $I\{p,\bar p\}$ satisfies, say
\be
\hat W_{-2} I = \bar p_2 I
\label{W-2}
\ee
with
\be
\hat W_{-2} := \sum_{a,b=0}^\infty \left( ab p_{a+b-2}\frac{\p^2}{\p p_a \p p_b}
+ (a+b+2)p_ap_b\frac{\p}{\p p_{a+b+2}} \right)
\ee
at $p_0=N$ -- what is indeed the case.
If restricted to Miwa-locus, $p_k=\sum_{i=1}^N z_i^k$ these relations become Calogero equations \cite{Cal},
they are easily lifted to an infinite system of commuting relations, describing IZ as a common eigenfunction of
commuting $W$-operators of \cite{MMMP,MMPdim}.
For brevity we call (\ref{W-2}) Calogero equation even on the entire space of time-variables.

Equation like (\ref{W-2}) is already easy to $\beta$-deform:
\be
\hat W^{(\beta)}_{-2} I^{(\beta)} =\frac{(\beta+1)c_{[1,1]}}{2c_{\emptyset}} \bar p_2 I^{(\beta)}
\label{Wbeta-2}
\ee
with
\be
\hat W^{(\beta)}_{-2} := \sum_{a,b=0}^\infty \left(  ab p_{a+b-2}\frac{\p^2}{\p p_a \p p_b}
+ \beta\cdot (a+b+2)p_ap_b\frac{\p}{\p p_{a+b+2}} \right) + (1-\beta)\sum_{a=0}^\infty (a+2)(a+1)p_a\frac{\p}{\p p_{a+2}}
\ee

What we need now is the check of (\ref{W-2}) and (\ref{Wbeta-2}) for the series expansion (\ref{Ichar}) and (\ref{Ibetachar}).

\bigskip

{\bf $\bullet$ Calogero equation for ordinary IZ:}
\be
I:=\sum_R \frac{S_R\{\delta_{k,1}\} S_R\{p\} S_R\{\bar p\}}{S_R[N]}
\ee
satisfies
\be
\hat W_{-2} I = \left(\sum_{i=1}^N y_i^2\right) I
\ee
with
\be
\hat W_{-2} := \sum_{a,b=0}^\infty \left( ab p_{a+b-2}\frac{\p^2}{\p p_a \p p_b}
+ (a+b+2)p_ap_b\frac{\p}{\p p_{a+b+2}} \right)
\ee
at $p_0=N$ and $\bar p_k = \tr Y^k = \sum_{i=1}^N y_i^k$.
\\

{\bf $\bullet$  $\beta$-deformed Calogero equation for the $\beta$-deformed IZ:}
\be
I^{(\beta)}:=\sum_R c_R\cdot \frac{J_R\{\delta_{k,1}\} J_R\{p\} J_R\{\bar p\}}{J_R[N]}
\label{Ibetachar1}
\ee
satisfies
\be
\hat W^{(\beta)}_{-2} I^{(\beta)} =\frac{(\beta+1)c_{[1,1]}}{2c_{\emptyset}} \left(\sum_{i=1}^N y_i^2\right) I^{(\beta)}
\ee
with
\be
\hat W^{(\beta)}_{-2} := \sum_{a,b=0}^\infty \left( \beta\cdot  ab p_{a+b-2}\frac{\p^2}{\p p_a \p p_b}
+ (a+b+2)p_ap_b\frac{\p}{\p p_{a+b+2}} \right) + (1-\beta)\sum_{a=0}^\infty (a+2)(a+1)p_a\frac{\p}{\p p_{a+2}}
\ee
and the weights
\be
c_{[2]} = \frac{(\beta+1)^2}{4\beta} \cdot c_{[1,1]}, \nn \\
\nn \\
c_{[3]} = \frac{(\beta+2)(\beta+1)^2}{12\beta^2} \cdot  \frac{c_{[1,1]} c_{[1]}}{c_{\emptyset}}, \nn \\
c_{[2,1]} = \frac{(\beta+1)(2\beta+1)}{2(\beta+2)\beta} \cdot   \frac{c_{[1,1]} c_{[1]}}{c_{\emptyset}}, \nn \\
c_{[1,1,1]} =  \frac{3}{2\beta+1}\cdot  \frac{c_{[1,1]} c_{[1]}}{c_{\emptyset}}, \nn \\
\nn \\
\ldots
\ee
Note that we do not include norms of Jack polynomials into the formula (\ref{Ibetachar1}),
like we did in  (\ref{Ibetachar}), and calculate the coefficients $c_R$ instead.
The ambiguity of normalization will be fixed in sec.\ref{mainbeta} below, see eq.(\ref{normc}):
\be
c_\emptyset = 1
\nn \\
c_{[1]}=\beta
\nn \\
c_{[2]} = \frac{(\beta+1)\beta}{2 } \ \ \Longleftarrow \ \
 c_{[1,1]} = \frac{2\beta^2}{\beta+1},
 \nn \\
 c_{[3]} =   \frac{(\beta+2)(\beta+1)\beta}{6}, \ \
 c_{[2,1]} = \frac{(2\beta+1)\beta^2}{\beta+2}, \ \
 c_{[1,1,1]} = \frac{6\beta^3}{(2\beta+1)(\beta+1)},
\label{cfrombetaCalogero}
\ee
{\footnotesize
\be
\!\!\!\!\!\!\!\!\!\!\!\!\!\! \!\!\!\!\!\!\!
c_{[4]} = \frac{(\beta+3)(\beta+2)(\beta+1)\beta}{24 }, \
c_{[3,1]} = \frac{(\beta+1)^2\beta^2}{\beta+3}, \
c_{[2,2]} = \frac{(2\beta+1)\beta^2}{\beta+2}, \
c_{[2,1,1]} = \frac{(3\beta+1)\beta^3}{(\beta+1)^2}, \
c_{[1,1,1,1]} = \frac{24\beta^4}{(3\beta+1)(2\beta+1)(\beta+1)},
\nn \\ \nn \\
\!\!\!\!\!\!\!\!\!\!\!\!\!\! \!\!\!\!\!\!\!\!\!\!\!\!\!\!
c_{[5]} = \frac{(\beta+4)(\beta+3)(]\beta+2)(\beta+1)\beta}{120 }, \
c_{[4,1]} = \frac{(2\beta+3)(\beta+2)(\beta+1)\beta^2}{6 (\beta+4)}, \
c_{[3,2]} = \frac{(2\beta+1)(\beta+1)^2\beta^3}{(\beta+3)(\beta+2)}, \
c_{[3,1,1]} = \frac{(3\beta+2)\beta^3}{2\beta+3}, \nn \\
\!\!\!\!\!\!\!\!\!\!\!\!\!\!
c_{[1,2,2]} = \frac{(3\beta+1)(2\beta+1)\beta^3}{(\beta+2)(\beta+1)^2}, \
c_{[1,1,1,2]} = \frac{6(4\beta+1)\beta^4}{(3\beta+2)(2\beta+1)(\beta+1)}, \
c_{[1,1,1,1,1]} = \frac{120\beta^5}{(4\beta+1)(3\beta+1)(2\beta+1)(\beta+1)},
\nn
\ee
}
\be
\ldots
\nn
\ee
As shown in \cite{MOP}, these are exactly the standard norms of Jack polynomials from \cite{MD}:
\be
c_{R}= ||J_R||^{-2}= \prod_{(i,j)\in R} \frac{\beta(-R_{j}^T +i-1)-R_i+j}{\beta(-R_{j}^T +i)-R_i+j-1}    
\label{Jacknorms}
\ee
for the standardly chosen Jack polynomials, like 
$J_{[3]}=\frac{\beta^2 p_1^3 + 3\beta p_2p_1 + 2p_3}{(\beta + 1)(\beta + 2)}$.
This makes   (\ref{Ibetachar}) exact without any extra coefficients,
what looks like the most natural deformation of (\ref{Ichar}).
This is exactly like it was with the $\beta$-deformation of
the superintegrability formulas \cite{SI,SIrev}.

\bigskip

{\bf $\bullet$ IZ as a common eigenfunction}

Another way to fix the normalization ambiguity is to impose other restrictions like (\ref{W-2}):
\be
\hat W_{-n}I:= \tr\!\! \left(\frac{\p}{\p X}\right)^n I = \left(\tr Y^n\right)\cdot  I
\label{W-n}
\ee
These are commuting Hamiltonians along the ray $(-1,1)$ in terms of \cite{MMMP}.
Like ordinary Schur polynomials are the common eigenfunctions of Hamiltonians along the ray $(0,1)$,
the {\bf IZ integral is the common eigenfunction for those along $(-1,1)$}
\footnote{This poses an interesting questions of how the Miki rotation \cite{Miki,MMMP}
can convert one into another (Schur into IZ),
and what are appropriate generalizations to all other rays.
The question becomes especially interesting after Macdonald deformation, inspired by \cite{MMPdim}.}.
This implies $\beta$-deformation of (\ref{W-n}).
The simplest complement of (\ref{Wbeta-2}) is then
\be
\hat W^{(\beta)}_{-1} I^{(\beta)} =\frac{ c_{[1]}}{ c_{\emptyset}} \cdot \bar p_1 I^{(\beta)}
\label{Wbeta-1}
\ee
with
\be
\hat W^{(\beta)}_{-1} :=  \sum_{a=0}^\infty  (a+1)p_a\frac{\p}{\p p_{a+1}}
\ee
Already this simple equation restricts $c_R$ to be
\be
c_{[2]} = \frac{\beta+1}{2\beta}\cdot \frac{c_{[1]}^2}{c_{\emptyset}}, \ \ \ \
c_{[1,1]} = \frac{2}{\beta+1} \cdot \frac{c_{[1]}^2}{c_{\emptyset}}, \nn \\
c_{[3]} = \frac{(\beta+2)(\beta+1)}{6\beta^2}\cdot \frac{c_{[1]}^3}{c_{\emptyset}^2}, \ \ \ \
c_{[2,1]} = \frac{(2\beta+1)}{\beta(\beta+2)}\cdot \frac{c_{[1]}^3}{c_{\emptyset}^2}, \ \ \ \
c_{[1,1,1]} = \frac{6 }{(2\beta+1)(\beta+1)}\cdot \frac{c_{[1]}^3}{c_{\emptyset}^2},
\nn \\
\ldots
\label{cs}
\ee
in accordance with (\ref{cfrombetaCalogero}) and (\ref{Jacknorms}).

\bigskip

{\bf $\bullet$ Macdonald case}

The same trick of using $\hat W_{-1}$ instead of $\hat W_{-2}$
can be also applied to study the further $(q,t)$-deformation of (\ref{Ichar})  with the help of \cite{MMPdim}.
From
\be
\hat W^{(q,t)}_{-1} I^{(q,t)} =-\frac{(1-q) C_{[1]}}{ (1-t^{-1}) C_{\emptyset}} \cdot \bar p_1 I^{(q,t)}
\label{Wqt-1}
\ee
with
\be
I^{(q,t)}[X,Y]:=\sum_{R: \ l_R\leq N}  C_R\cdot \frac{M_R\{\delta^*_{k,1}\} M_R[X] M_R[Y]}{  M_R[N^{^*}]}
\label{Iqtchar1}
\ee
and
\be
\hat W^{(q,t)}_{-1} := {\rm Reg}\left\{
-\frac{1}{(1-t^{-1})(1-q)}\oint_0 dz\ {\rm exp}\left(\sum_{k=1}\frac{1-t^{-k}}{k}z^kp_k\right)
{\rm exp}\left(-\sum_{k=1}\frac{1-q^{k}}{z^k}\frac{\p}{\p p_k}\right)\right\} = \nn \\
= \underline{\frac{1-t^{-N}}{1-t^{-1}}}\frac{\p}{\p p_1}
+ p_1\left((q+1)\frac{\p}{\p p_2} + \frac{q-1}{2}\frac{\p^2}{\p p_1^2}\right) + \nn \\
+ \frac{(1+t^{-1})p_2+(1-t^{-1})p_1^2}{2}\left((1+q+q^2)\frac{\p}{\p p_3} + (q^2-1)\frac{\p^2}{\p p_2\p p_1}
+ \frac{(q-1)^2}{6}\frac{\p^3}{\p p_1^3} \right) + \ldots
\label{Wqt-1}
\ee
it follows that
\be
C_{[2]} = \frac{qt-1}{(t-1)(q+1)}\cdot \frac{C_{[1]}^2}{C_{\emptyset}}, \ \ \ \
C_{[1,1]} = \frac{(q-1)(t+1)}{qt-1} \cdot \frac{C_{[1]}^2}{C_{\emptyset}},
\label{Cs}
\ee
{\footnotesize
\be
C_{[3]} = \frac{(q^2t-1)(qt-1)}{(q^2+q+1)(q+1)(t-1)^2}\cdot \frac{C_{[1]}^3}{C_{\emptyset}^2}, \ \ \ \
C_{[2,1]} = \frac{(qt^2-1)(q-1)}{(q^2t-1)(t-1)}\cdot \frac{C_{[1]}^3}{C_{\emptyset}^2}, \ \ \ \
C_{[1,1,1]} = \frac{(t^2+t+1)(t+1)(q-1)^2 }{(qt-1)(qt^2-1)}\cdot \frac{C_{[1]}^3}{C_{\emptyset}^2},
\nn
\ee
}
\be
\ldots
\nn
\ee
Like in (\ref{Jacknorms}),
\be
    C_{R}\sim ||M_R||^{-2} =\prod_{k=1}^{k=N} \frac{1-q^{a(k)+1}t^{l(k)}}{1-q^{a(k)}t^{l(k)+1}}
\label{MDnorms}
\ee
where $a(k)$ and $l(k)$ are the lengths of $k$'s hook arm and leg respectively \cite{MD}. 
This normalization refers to the standard choice of Macdonalds, like 
$ M_{[2]} = \frac{(q+1)(t-1)p_1^2+(q-1)(t+1)p_2}{2(qt-1)}$,
and the proportionality refers to the freedom in selection of $C_{\emptyset}$ and $C_{[1]}$,
which remains when we do not take into account the other operators $\hat W^{(q,t)}_m$. 

Regularization operation ``Reg" in the first line in (\ref{Wqt-1}) refers to modification of the underlined
coefficient in the second line, which otherwise would be singular in the limit $t\longrightarrow 1$.
In the limit $\hbar\longrightarrow 0$ with $q=e^\hbar$, $t=e^{\beta\hbar}$ the constants (\ref{Cs}) reproduce
(\ref{cs}).
They are also related to the standard normalization of Macdonald polynomials.
Of course more detailed consideration is needed of $\hat W^{(q,t)}_{-2}$ and higher commuting Hamiltonians.
They are given by multiple $z$-integrals \cite{MMPdim} and are  more tedious to analyze.
Still their commutativity implies that the results will not actually change too much.

\section{Gaussian averages }

The next point is a kind of sum rules, implied by decomposition (\ref{Ichar}).
Namely,
\be
\begin{array}{ccccc}
\int I[X,Y]e^{-\frac{1}{2}\tr Y^2} dY && \stackrel{(\ref{Ichar})}{=} &&
\sum_R \frac{S_R\{\delta_{k,1}\} S_R[X]}{S_R[N]} \int S_R[Y] e^{-\frac{1}{2}\tr Y^2} \\
\\
|| &&&& || \ SI  \\
\\
e^{-\frac{1}{2}\tr X^2} & \stackrel{\rm Cauchy}{=} &
\sum_R S_R\{\delta_{k,2}\}S_R[X] &=& \sum_R  \frac{S_R\{\delta_{k,1}\} S_R[X]}{S_R[N]}
\frac{S_R\{\delta_{k,2}\} S_R[N]}{S_R\{\delta_{k,1}\}}
\end{array}
\ee
where the right vertical identity follows from the superintegrability (SI) property \cite{SI,SIrev}
of Gaussian integrals, while the left identity in the bottom line is the Cauchy formula \cite{Cauchy}
for the Schur functions.

$\beta$-deformation is straightforward.

More interesting is the generalization of superintegrability to pair correlator \cite{bilin},
which proved to be very useful in the WLZZ theory, see sec.4 of \cite{MMPSh2}:
\be
\left<K_Q S_R\right>_G =  \left<\hat W_Q^- S_R\right>_G
= \left<\prod_{i=1}^{l_Q}\tr \left(\frac{\p}{\p Y}\right)^{Q_i} S_R\right>_G
= \eta_R(N)S_{R/Q}\{\delta_{k,2}\}
\ee
and
\be
\left<K_Q K_R\right>_G = \eta_R(N) \delta_{R,Q}
\ee
where $\eta_R(N):=\frac{S_R[N]}{S_R\{\delta_{k,1}\}}$.

\section{Regularization of integrals and their series expansions}

However, the Gaussian averages from the previous section can not be the end of the story.
There are important integrals with the IZ weight only.
The typical example is orthogonality relation
\be
\int S_R[X] S_Q[Y] e^{i\tr XY} dX dY \sim \delta_{R,Q}
\ee

{\bf $\bullet$ Orthogonality w.r.t. the IZ measure:}

It can be deduced directly from its elementary version at $N=1$:
\be
\int\int x^m y^n e^{ixy} dxdy = \frac{2\pi}{i^n} \int x^m\delta^{(n)}(x) dx = 2\pi i^n\cdot  n! \cdot \delta_{n,m}
\label{orthopol}
\ee
The brute force calculation leads to
\be
\int S_R[X] S_Q[Y] e^{i\tr XY} dX dY =
\delta_{Q,R} \cdot \frac{S_R[N]}{S_R\{\delta_{k,1}\}}\cdot
 \underbrace{\int e^{i\tr XY} dX dY}_{\prod_{m=1}^N 2\pi i^m\cdot  m!}
\label{orthochar}
\ee
It is based on the substitution $e^{i\tr XY} dX dY \sim  \Delta[X]\Delta[Y] \prod_{j=1}^N e^{i x_jy_j}dx_jdy_j$
for averaging of the time-dependent functions $F\{\tr X^k,\tr Y^l\}$, i.e. depending only on the symmetric combinations
of the eigenvalues.

\bigskip

{\bf $\bullet$ Orthogonality from the expansion of (\ref{orthopol})}.
However, for $\beta$-deformation this matrix trick does not work
and we are going to use character representation of IZ function.
Thus what we need is to understand how (\ref{orthopol}) works, when the exponential  is expanded and integrals diverge.
To make this possible, we need to regularize the integral, e.g. by introducing a Gaussian weight.
Then we need to prove that
\be
\lim_{\alpha\rightarrow +0}
\int\int x^n y^m \sum_{k=0}^\infty \frac{(ixy)^k}{k!} e^{-\alpha x^2-\alpha y^2} dxdy
= 2\pi i^n\cdot n!\cdot \delta_{n,m}
\ee

To begin with, put $n=m=0$, then
\be
\sum_{k=0}^\infty \int\int  \frac{(ixy)^{2k}}{(2k)!} e^{-\alpha x^2-\alpha y^2} dxdy =
\frac{ \pi}{\alpha}\sum_{k=0}^\infty \frac{(2k )!}{(k!)^2 } (-16\alpha^2)^{-k}
= \frac{\pi}{\alpha}\frac{1}{\sqrt{1+\frac{1}{4\alpha^2}}}
= \frac{2\pi}{\sqrt{1+4\alpha^2}}
\ee
Despite each term in the expansion is singular in the limit $\alpha\rightarrow 0$,
and the series is formally divergent for $4\alpha^2<1$,
the analytical continuation and
the limit of the {\it sum} is well defined and reproduces the desired (\ref{orthopol}).

Likewise
\be
\boxed{
\int\int (xy)^{2n} \sum_{k=0}^\infty \frac{(ixy)^{2k}}{(2k)!} e^{-\alpha x^2-\alpha y^2} \frac{dxdy}{2\pi}
= \frac{ (-)^n (2n)!}{(1+4\alpha^2)^{n+\frac{1}{2}}}\cdot\left( 1 -\frac{2n(2n+1)\alpha^2}{1+4\alpha^2}
+ \ldots       \right) \ \stackrel{\alpha\rightarrow +0}{\longrightarrow} \ (-)^n (2n)!
}
\ee
Examples of correction terms, irrelevant in the limit $\alpha\rightarrow +0$:
\be
\begin{array}{c|l}
2n=0 & 1 \\
2n=2 & 1 - \frac{2\cdot 3\cdot \alpha^2}{1+4\alpha^2} \nn \\
2n=4 & 1 - \frac{4\cdot 5\cdot \alpha^2}{1+4\alpha^2} + \frac{2\cdot 5\cdot 7\cdot  \alpha^4}{(1+4\alpha^2)^2} \nn \\
2n=6 & 1 - \frac{6\cdot 7\cdot \alpha^2}{1+4\alpha^2} + \frac{6\cdot 7\cdot 9\cdot  \alpha^4}{(1+4\alpha^2)^2}
-  \frac{12\cdot 7\cdot 11\cdot  \alpha^6}{(1+4\alpha^2)^3} \nn \\
2n=8 & 1 - \frac{8\cdot 9\cdot \alpha^2}{1+4\alpha^2} + \frac{12\cdot 9\cdot 11 \cdot  \alpha^4}{(1+4\alpha^2)^2}
-  \frac{6\cdot 8\cdot 11\cdot 13\cdot  \alpha^6}{(1+4\alpha^2)^3}
+  \frac{6\cdot 11\cdot 13\cdot 15\cdot  \alpha^8}{(1+4\alpha^2)^4} \nn \\
\ldots
\end{array}
\ee
and
\be
\!\!\!\!\!\!\!\!\!\!\!
\boxed{
\int\int (xy)^{2n+1} \sum_{k=0}^\infty \frac{(ixy)^{2k+1}}{(2k+1)!} e^{-\alpha x^2-\alpha y^2} \frac{dxdy}{2\pi}
= \frac{ i\cdot(-)^n (2n+1)!}{(1+4\alpha^2)^{n+\frac{1}{2}}}\cdot\left( 1 -\frac{2n(2n+3)\alpha^2}{1+4\alpha^2}
+ \ldots       \right) \ \stackrel{\alpha\rightarrow +0}{\longrightarrow} \ i\cdot (-)^n (2n+1)!
}
\nn
\ee
\be
\begin{array}{c|l}
2n+1=1 & 1 \\
2n+1=3 & 1 - \frac{2\cdot 5\cdot \alpha^2}{1+4\alpha^2} \nn \\
2n+1=5 & 1 - \frac{4\cdot 7\cdot \alpha^2}{1+4\alpha^2} + \frac{2\cdot 7\cdot 9\cdot  \alpha^4}{(1+4\alpha^2)^2} \nn \\
2n+1=7 & 1 - \frac{6\cdot 9\cdot \alpha^2}{1+4\alpha^2} + \frac{6\cdot 9\cdot 11\cdot  \alpha^4}{(1+4\alpha^2)^2}
-  \frac{11\cdot 12\cdot 13\cdot  \alpha^6}{(1+4\alpha^2)^3} \nn \\
2n+1=9 & 1 - \frac{8\cdot 11\cdot \alpha^2}{1+4\alpha^2} + \frac{11\cdot 12\cdot 13 \cdot  \alpha^4}{(1+4\alpha^2)^2}
-  \frac{80\cdot 11\cdot 13\cdot  \alpha^6}{(1+4\alpha^2)^3}
+  \frac{10\cdot 11\cdot 13\cdot 17\cdot  \alpha^8}{(1+4\alpha^2)^4} \nn \\
\ldots
\end{array}
\ee
As an illustration of orthogonality we mention just
\be
\!\!\!\!\!\!\!\!\!\!\!
\boxed{
\int\int x^{2n} \sum_{k=0}^\infty \frac{(ixy)^{2k}}{(2k)!} e^{-\alpha x^2-\alpha y^2} \frac{dxdy}{2\pi}
= \frac{ 2^n(2n-1)!!\cdot \alpha^n}{(1+4\alpha^2)^{n+\frac{1}{2}}}  \ \stackrel{\alpha\rightarrow +0}{\longrightarrow} \ \delta_{n,0}
}
\ee
In this particular case there are no corrections.

\bigskip

{\bf $\bullet$ Another avatar of Gaussian regularization:  the limit of pair correlator.}
Let
\be
\Big<F\Big>_{\alpha}:=\frac{1}{(2\pi)^{N^2/2}} \int F(X) e^{-\frac{\alpha}{2}\tr X^2} dX
\ee
Then superintegrability \cite{SI,SIrev} implies
\be
\left<S_R[X]\right>_{\alpha} = \frac{1}{\alpha^{\frac{|R|+N^2}{2}}}\cdot
\frac{S_R\{\delta_{k,2}\}S_R[N]}{S_R\{\delta_{k,1}\}}
\ee
At the same  time we can split Gaussian exponential in two and apply Cauchy formula \cite{Cauchy}
\be
e^{\sum_k \frac{p_k\bar p_k}{k}} = \sum_Q S_R\{p\}S_R\{\bar p\} \ \ \ \Longrightarrow \ \ \ \
e^{\frac{\mu}{2}\tr X^2} = \sum_Q \mu^{|Q|/2}S_Q\{\delta_{k,2}\}S_Q[X]
\ee
to the remaining piece:
\be
\Big<S_R[X]\Big>_{\alpha-\mu} =
\Big<S_R[X] e^{\frac{\mu}{2}\tr X^2}\Big>_\alpha =
\sum_Q \mu^{|Q|/2}\cdot S_Q\{\delta_{k,2}\} \Big<S_Q[X]S_R[X]\Big>_\alpha
\ee
In particular. for $R=\emptyset$
\be
\!\!\!\!\!\!\!\!\!\!\!
\Big<1\Big>_{\alpha-\mu} = \frac{1}{(\alpha-\mu)^{N^2/2}} = \sum_Q \mu^{|Q|/2}\cdot S_Q\{\delta_{k,2}\} \Big<S_Q[X] \Big>_\alpha
= \frac{1}{(\alpha)^{N^2/2}}\sum_Q \left(\frac{\mu}{\alpha}\right)^{|Q|/2}\cdot
\frac{ \left(S_Q\{\delta_{k,2}\}\right)^2S_Q[N]}{S_Q\{\delta_{k,1}\}}
\ee
Thus, despite the r.h.s. consists of terms with $\alpha$ in denominators, the l.h.s. implies that this series
can be explicitly summed up and the sum makes sense even when $\alpha\longrightarrow 0$.
Moreover, these summation tricks can be promoted to the case of pair correlators,
which are not defined from superintegrability property --
i.e. the summation is possible irrespective of superintegrability.

\section{The character expansion of the 2-matrix model}

Now we apply the experience from the previous section  to the case of the double-matrix model,
which is usually analyzed with the help of eq.(\ref{Iint}).
However, our purpose is to develop a $\beta$-deformable approach.
Therefore we apply eq.(\ref{Ichar}) instead.

\bigskip

{\bf $\bullet$ The main  relation} which we wish to deduce is
\be
\int\int e^{\sum_k \frac{1}{k}\left(p_k\tr X^k +\bar p_k\tr Y_k\right)} e^{\tr XY}dXdY
\ \stackrel{?}{=}\ \sum_R \frac{S_R[N]S_R\{p\}S_R\{\bar p\}}{S_R\{\delta_{k,1}\}}
\label{mainrel}
\ee
and at the l.h.s. we want to substitute  the character expansion through the eigenvalues:
\be
e^{\tr XY}\  \Longrightarrow \  I(X,Y) = \sum_R \frac{S_R\{\delta_{k,1}\} S_R[X] S_R[Y] }{S_R[N]}
\ee
We omit $i$ from the exponential to simplify the majority of formulas.
And we put question sign in (\ref{mainrel}) to emphasize that we are in the process of derivation.

The integrals over $X$ and $Y$ at the l.h.s. of (\ref{mainrel}) can  be reduced to Gaussian, provided
we shift the time-variables $p_k \longrightarrow  -\mu\delta_{k,2} + p_k$,
$\bar p_k \longrightarrow  -\nu\delta_{k,2} + \bar p_k$, and expand around the Gaussian point $p=\bar p=0$.
The remaining averages will be of Schur functions and can be handled by superintegrability formulas.

Our immediate interest is, however, different.
The point is that the l.h.s. is expanded in {\it negative} powers of $\mu$, $\nu$.
At the same time the r.h.s.  is  naturally expanded in {\it positive} powers of $\mu,\nu$.
Thus we convert (\ref{mainrel}) to the following puzzling form:
\be
\boxed{
(-)^{N^2/2}
\sum_R \frac{S_R[N] S_R\{\delta_{k,2}\}^2}{(\mu\nu)^{\frac{N^2+|R|}{2}}S_R\{\delta_{k,1}\}} + \ldots
\stackrel{?}{=}\ \sum_R  (\mu\nu)^{\frac{|R|}{2}}\frac{S_R[N] S_R\{\delta_{k,2}\}^2}{S_R\{\delta_{k,1}\}} + \ldots
}
\label{mainSchurp=0}
\ee
where $\ldots$ denote $p$-dependent corrections, which we neglect for a while,
till (\ref{relwithLR}) below.
The reason for (\ref{mainSchurp=0}) to hold is that
both sides are proportional to $\frac{1}{(1-\mu\nu)^{N^2/2}}$ (!) and can be easily expanded in
both positive and negative powers:
\be
\frac{(-)^{N^2/2}}{(\mu\nu)^{N^2/2}} \sum_m \frac{\xi_m}{(\mu\nu)^m}
= \frac{1}{(1-\mu\nu)^{N^2/2}} =
\sum_m \xi_m\cdot (\mu\nu)^m
\ee
with the same coefficients $\xi_m = \sum_{|R|=2m} \frac{S_R[N] S_R\{\delta_{k,2}\}^2}{ S_R\{\delta_{k,1}\}} $ on both sides.
{\bf For things to work this way,
expanded should be just powers of $(1-\mu\nu)$ --
what is the case here and will remain the case after the $\beta$-deformation, see (\ref{mainJackp=0}).
}

The identity behind (\ref{mainSchurp=0}) are as follows: the elementary
\be
\int e^{- \frac{\mu}{2} \tr X^2 - \frac{\nu}{2} \tr Y^2 } e^{ \tr XY} dX dY
= {\rm det}^{-\frac{N^2}{2}}\left(\begin{array}{cc} \mu & -1 \\ -1 & \nu \end{array}\right)
= (\mu\nu-1)^{-\frac{N^2}{2}}
\ee
where we now assume the convenient normalization $\int e^{i\tr XY} dXdY=1$,
and a more transcendental equality
\be
\sum_Q    \frac{S_Q[N]S_Q\{\delta_{k,2}\}^2} {S_Q\{\delta_{k,1}\}}\cdot (\mu\nu)^{|Q|/2 }
= (1-\mu\nu)^{-\frac{N^2}{2}}
\label{ppIZ2}
\ee
which one can check by  direct calculation.

\bigskip
{\bf $\bullet$ Corrections} to (\ref{mainSchurp=0}), coming from $p$ and $\bar p$ dependencies, are tedious, still straightforward.
In more detail, the l.h.s of (\ref{mainrel}) is
\be
\sum_R \frac{S\{\delta_{k,1}\}}{S_R[N]}  \sum_{R_1,R_2} S_{R_1}\{p\} S_{R_2}\{\bar p\}
\Big<S_R[X]S_{R_1}[X]\Big>_\mu  \Big< S_R[Y]S_{R_2}[Y]\Big>_\nu
= \nn \\
= \sum_R \frac{S\{\delta_{k,1}\}}{S_R[N]}  \sum_{R_1,R_2}  S_{R_1}\{p\} S_{R_2}\{\bar p\}
\sum_{Q_1}N^{Q_1}_{RR_1} \frac{S_{Q_1}\{\delta_{k,2}\}S_{Q_1}[N]}{\mu^{\frac{N^2+|Q_1|}{2}}S_{Q_1}\{\delta_{k,1}\}}
\sum_{Q_2}N^{Q_2}_{RR_2} \frac{S_{Q_2}\{\delta_{k,2}\}S_{Q_2}[N]}{\nu^{\frac{N^2+|Q_2|}{2}}S_{Q_2}\{\delta_{k,1}\}}
\label{relwithLR}
\ee
where  we substituted
\be
 \Big<S_R[X]S_{R_1}[X]\Big>_\mu =
\sum_{Q_1}N^{Q_1}_{RR_1} \Big<S_{Q_1}[X]\Big>_\mu =
\sum_{Q_1}N^{Q_1}_{RR_1}
\frac{S_{Q_1}\{\delta_{k,2}\}S_{Q_1}[N]}{\mu^{\frac{N^2+|Q_1|}{2}}S_{Q_1}\{\delta_{k,1}\}}
\ee
and a similar expression in the $Y$ sector.

\bigskip

The r.h.s. of (\ref{mainrel}) is
\be
 \sum_R \frac{S_R[N]S_R\{-\mu\delta_{k,2}+p_k\}S_R\{-\nu\delta_{k,2}+\bar p_k\}}{S_R\{\delta_{k,1}\}}
= \nn \\
= \sum_R    \frac{S_R[N]}{S_R\{\delta_{k,1}\}}
\sum_{R_1}  S_{R_1}\{p\} S_{R/R_1}\{-\mu\delta_{k,2}\}
\sum_{R_2} S_{R_2}\{\bar p_k\} S_{R/R_2}\{-\nu\delta_{k,2}\}
= \nn \\
= \sum_R    \frac{S_R[N]}{S_R\{\delta_{k,1}\}}
\sum_{R_1,Q_1}\mu^{|Q_1|/2}  N^R_{R_1Q_1} S_{R_1}\{p\} S_{Q_1}\{\delta_{k,2}\}
\sum_{R_2,Q_2} \nu^{|Q_2|/2} S_{R_2}\{\bar p\} S_{Q_2}\{\delta_{k,2}\}
= \nn \\
= \sum_R    \frac{S_R[N]}{S_R\{\delta_{k,1}\}} \sum_{R_1,R_2} S_{R_1}\{p\}  S_{R_2}\{\bar p\}
\sum_{Q_1}\mu^{|Q_1|/2}  N^R_{R_1Q_1} S_{Q_1}\{\delta_{k,2}\}
\sum_{Q_2} \nu^{|Q_2|/2}  S_{Q_2}\{\delta_{k,2}\}
 \ee
Alternatively, the r.h.s. of (\ref{mainrel}) can be rewritten as
\be
 \sum_R \frac{S_R[N]S_R\{-\mu\delta_{k,2}+p_k\}S_R\{-\nu\delta_{k,2}+\bar p_k\}}{S_R\{\delta_{k,1}\}}
= \nn \\
= \frac{1}{(1-\mu\nu)^{N^2/2}}\left(1
- \frac{N^2(\nu p_2+\mu \bar p_2)+ N(\nu p_1^2 - 2p_1\bar p_1 +\mu \bar p_1^2)}{2(1-\mu\nu)}
+ \ldots   \right)
\label{rhsmainrel}
\ee
while the l.h.s. of (\ref{mainrel}) begins from the contribution of $R_1=R_2=\emptyset$:
\be
\sum_R \frac{S_R\{\delta_{k,1}\}}{S_R[N]} \int\int S_{R[X]}S_R[Y]
e^{\sum_k \frac{1}{k}\left(p_k \tr X^k + \bar p_k \tr Y_k\right)}
e^{-\frac{\mu}{2}\tr X^2 - \frac{\nu}{2}\tr Y^2} dXdY
= \nn \\
= \sum_R \frac{S_R\{\delta_{k,1}\}}{S_R[N]}
\left(\sum_{R_1}S_{R_1}\{p\}  \Big<S_R[X]S_{R_1}[X]\Big>_\mu\right)
\left(\sum_{R_2}S_{R_2}\{\bar p\} \Big<S_R[Y]S_{R_2}[Y]\Big>_\nu\right)
= \nn\\
= \sum_R \frac{S_R\{\delta_{k,1}\}}{S_R[N]}
\cdot \left(\frac{S_R\{\delta_{k,2}\}S_R[N]}{\mu^{\frac{N^2+|R|}{2}}\cdot S_R\{\delta_{k,1}\}} + O(p)\right)
\cdot \left(\frac{S_R\{\delta_{k,2}\}S_R[N]}{\nu^{\frac{N^2+|R|}{2}}\cdot S_R\{\delta_{k,1}\}} + O(\bar p)\right)
= \nn \\
= \frac{1}{(\mu\nu)^{N^2/2}}\left(1-\frac{1}{\mu\nu}\right)^{-N^2/2}\Big(1 + O(p,\bar p)\Big)
\ee
The $p,\bar p$-dependent corrections are sensitive to the Littlewood-Richardson coefficients $N_{RR_1}^{Q_1}$.
The first relevant products of representations (those of even size) are :
\be
\!\!\!\!\!\!\!\!
\begin{array}{c||c||c||c|c||c|c|c||cc}
R_1\backslash R & [0] & [1] & [2] & [11] & [3] & [21] & [111] & \ldots \\
\hline
\phantom.[0] &   [0]  &  - & [2] & [11] &- & -& - & \ldots \\
\phantom.[1] &  -  &  [2]+[11] & - & - & [4]+[31]&  [31]+[22]+[211] & [211]+[1111] & \ldots \\
\phantom.[2] &   [2]  &  - &  [4]+[31]+[22] & [31]+[211] & - & - & - & \ldots \\
\phantom.[11] &  [11] & - & [31]+[211] & [22]+[211]+[1111] & - & - & -   & \ldots \\
\ldots
\end{array}
\nn
\ee

\be
\sum_R \frac{S_R\{\delta_{k,1}\}}{S_R[N]} \int\int S_R[X]S_R[Y]
e^{\sum_k \frac{1}{k}\left(p_k \tr X^k + \bar p_k \tr Y_k\right)}
e^{-\frac{1}{2u^2}\tr X^2 - \frac{1}{2v^2}\tr Y^2} dXdY
= \nn \\
= \sum_R \frac{S_R\{\delta_{k,1}\}}{S_R[N]}
\left(\sum_{R_1}S_{R_1}\{p\}  \Big<S_R[X]S_{R_1}[X]\Big>_{\frac{1}{u^2}}\right)
\left(\sum_{R_2}S_{R_2}\{\bar p\} \Big<S_R[Y]S_{R_2}[Y]\Big>_{\frac{1}{v^2}}\right)
= \nn\\
= (uv)^{N^2}\sum_R \frac{S_R\{\delta_{k,1}\}}{S_R[N]}
\cdot \left(u^{|R|}\frac{S_R\{\delta_{k,2}\}S_R[N]}{ S_R\{\delta_{k,1}\}} + O(p)\right)
\cdot  \left(v^{|R|}\frac{S_R\{\delta_{k,2}\}S_R[N]}{   S_R\{\delta_{k,1}\}} + O(\bar p)\right)
= \nn \\
= \frac{(uv)^{N^2}}{(1-u^2v^2)^{N^2/2}}\Big(1 + O(p,\bar p)\Big)
\ee
Introduce $G_{R\big|R_1R_2}:= \frac{S_R\{\delta_{k,1}\}}{S_R[N]} S_{R_1}\{p\}S_{R_2}\{\bar p\}$
and pull our the factor of $u$ and $v$ from the averages.
Then the first items in this sum are
{\footnotesize
\be
(uv)^{N^2}\Big\{
G_{[0]\big|[0],[0]} \cdot \Big<S_{[0]}\Big>^2
+ \nn \\
{\bf u^2}\left\{G_{[0]\big|[2],[0]} \cdot  \Big<S_{[2]}\Big>\Big<S_{[0]}\Big>
+ G_{[0]\big|[1,1],[0]} \cdot  \Big<S_{[1,1]}\Big>\Big<S_{[0]}\Big>\right\}
+ \nn \\
+{\bf v^2}\left\{ G_{[0]\big|[0],[2]} \cdot   \Big<S_{[0]}\Big> \Big<S_{[2]}\Big>
 + G_{[0]\big|[0],[1,1]} \cdot  \Big<S_{[0]}\Big> \Big(S_{[1,1]}\Big>\right\}
+ \nn \\
+ {\bf u^4v^2}\left\{ G_{[2]\big|[2],[0]} \cdot
\Big(\left<S_{[4]}\right>+\left<S_{[3,1]}\right>+\left<S_{[2,2]}\right>\Big)\left<S_{[2]}\right>
+ G_{[2]\big|[1,1],[0]} \cdot
\Big( \left<S_{[3,1]}\right>+\left<S_{[2,1,1]}\right>\Big)\left<S_{[1,1]}\right>
+ \right. \nn \\ \left.
+ G_{[1,1]\big|[2],[0]} \cdot
\Big(\left<S_{[3,1]}\right>+\left<S_{[2,1,1]}\right>\Big)\left<S_{[2]}\right>
+ G_{[1,1]\big|[1,1],[0]} \cdot
\Big(\left<S_{[2,2]}\right>+\left<S_{[2,1,1]}\right>+\left<S_{[1,1,1,1]}\right>\Big)\left<S_{[1,1]}\right>
\right\}
+ \nn \\
+ {\bf u^2v^4}\left\{ G_{[2]\big|[0],[2]} \cdot
\left<S_{[2]}\right>\Big(\left<S_{[4]}\right>+\left<S_{[3,1]}\right>+\left<S_{[2,2]}\right>\Big)
+ G_{[2]\big|[0],[1,1]} \cdot
\left<S_{[1,1]}\right>\Big( \left<S_{[3,1]}\right>+\left<S_{[2,1,1]}\right>\Big)
+ \right. \nn \\ \left.
+ G_{[1,1]\big|[0],[2]} \cdot
\left<S_{[2]}\right>\Big(\left<S_{[3,1]}\right>+\left<S_{[2,1,1]}\right>\Big)
+ G_{[1,1]\big|[0],[1,1]} \cdot
\left<S_{[1,1]}\right>\Big(\left<S_{[2,2]}\right>+\left<S_{[2,1,1]}\right>+\left<S_{[1,1,1,1]}\right>\Big)
\right\}
+ \nn \\
\!\!\!\!\!\!\!\!\!\!\!\!\!\!\!\!
+ {\bf u^4v^4}\left\{G_{[3]\big|[1],[1,1]}   \Big(\!\left<S_{[4]}\right>+\left<S_{[3,1]}\right>\!\Big)^2
+ G_{[2,1]\big|[1],[1,1]}
\Big(\!\left<S_{[3,1]}\right>+\left<S_{[2,2]}\right>+\left<S_{[2,1,1]}\right>\!\Big)^2
+ G_{[1,1,1]\big|[1],[1,1]} \Big(\!\left<S_{[2,1,1]}\right>+\left<S_{[1,1,1,1]}\right>\!\Big)^2
\right\}
+ \nn \\
+ \ldots \Big\} =
\nn
\ee
}
\be
=\frac{(uv)^{N^2}}{(1-u^2v^2)^{N^2/2}}
\left(1 + \frac{N^2(u^2p_2+v^2\bar p_2) + N(u^2p_1^2+2u^2v^2p_1\bar p_1 + v^2\bar p_1^2)}{2(1-u^2v^2 )}
+ \ldots \right)
\label{lhsmainrel}
\ee

\bigskip

{\bf $\bullet$ $\beta$-deformation \label{mainbeta}}

The $\beta$-deformed analogue of (\ref{mainSchurp=0}) is
\be
\boxed{
\sum_R c_R\cdot \frac{J_R[N] J_R\{\delta_{k,2}\}^2}{(\mu\nu)^{\frac{N+\beta N(N-1)+|R|}{2}}J_R\{\delta_{k,1}\}}
= \sum_R  (\mu\nu)^{\frac{|R|}{2}}c_R \cdot \frac{J_R[N] J_R\{\delta_{k,2}\}^2}{J_R\{\delta_{k,1}\}}
= \frac{1}{(1-\mu\nu)^{\frac{N+\beta N(N-1)}{2}}}
}
\label{mainJackp=0}
\ee
with $c_R$ equal to
\be
c_{[0]}=1, \nn \\
c_{[2]} = \frac{\beta(\beta+1)}{2}, \ \ \ c_{[1,1]} = \frac{2\beta^2}{\beta+1}, \nn \\
 c_{[3]} =   \frac{(\beta+2)(\beta+1)\beta}{6}, \ \
 c_{[2,1]} = \frac{(2\beta+1)\beta^2}{\beta+2}, \ \
 c_{[1,1,1]} = \frac{6\beta^3}{(2\beta+1)(\beta+1)}, \nn \\
\ldots
\label{normc}
\ee
in full agreement with (\ref{cfrombetaCalogero}) and (\ref{Jacknorms}).
Note that the duality relation (\ref{mainJackp=0}) holds only for this very special choice of $c_R$.
Only diagrams of even size contribute to (\ref{mainJackp=0}) because of the factors $J_R\{\delta_{k,2}\}$.
Moreover, $c_{[2,2]}$ is not defined unambiguously by the requirement (\ref{mainJackp=0}),
we take it from (\ref{cfrombetaCalogero}).
Both these facts mean that (\ref{mainJackp=0}) is {\it consistent}
with (\ref{cfrombetaCalogero}), but does not {\it imply} it.
In exchange, the (normalization) freedom in (\ref{cfrombetaCalogero}) is restricted/fixed by (\ref{mainJackp=0}).

\section{Conclusion}

In this paper we demonstrated regularization tricks to deal with the character expansion of IZ integrals.
Besides emphasising the puzzling general problem of interrelation between integral and series solutions
of Ward identities,
this representation allows straightforward  deformations, like Jack ($\beta$) and Macdonald ($q,t$),
where clever integral formulas are not yet available.
The price to pay is complication with  further integration over eigenvalues,
even Gaussian, which is of crucial importance for  applications.
This contradiction was skillfully resolved recently in \cite{MOP},
what allowed a thorough study and $\beta$-deformation of integral 2-matrix realization \cite{MMMP} of
the WLZZ models \cite{WLZZ}, which were originally defined via the $W$-representations
\cite{MS,WrepMM}
and lacked a matrix-model realization.
The present paper describes a somewhat different approach, explaining the way to regularize
and sum the divergent series, formally associated with $\delta$-function integrals,
which stand behind the orthogonality properties of characters.
The WLZZ models \cite{WLZZ} made it clear that these properties can imply much more than expected
and one can use them for a deeper study of Yangian and DIM symmetries \cite{YDIM}
with the help of the matrix model technique.
The Gaussian regularization can play a big role in this approach.
It is also interesting by itself, even some of the identities in this paper require better understanding
and transparent explanations.
Also one can now revisit the problem of IZ correlators \cite{MIZcorr},
perhaps, with the help of methods, outlined in Appendix B of \cite{MMMPcorr}.
A separate story is ($q,t$)-deformation, which is only touched at the end of sec.\ref{Calo} of this paper
and deserves  more detailed consideration.
An interesting question here is also about the Miki rotation \cite{Miki,MMMP},
 relating common eigenfunctions for different
integrable subsystems (rays), which should convert Macdonald polynomials for the vertical ray $(0,1)$
into the IZ series (\ref{Iqtchar}) for the $(-1,1)$ one.
Since coefficients $c_R$ and $C_R$ appear related to the standard norms of Jack/Macdonald polynomials,
which appear also in the superintegrability formulas \cite{SI,SIrev},
a long-standing problem of further generalization to Kerov functions \cite{Kerov} is also getting a new momentum.

\bigskip

To summarize,  this paper opens at least five directions for further research:
\begin{itemize}
\item{Duality relations like (\ref{mainSchurp=0}) and (\ref{mainJackp=0}) for series,
convergent in complementary domains, which possess a unifying analytical continuation --
and their connection to complementarity  between integral and series solutions of the Ward identities}

\item{Macdonald generalization (\ref{Iqtchar}) of the IZ expansion
and possible extension to Kerov functions}

\item{The role of IZ as the common eigenfunction for the $(q,t)$-deformation of Calogero Hamiltonians
${\rm tr}\left(\frac{\p}{\p X}\right)^n$, i.e. ray $(-1,1)$ --
and its Miki-like relation to Schur-Macdonald polynomials, which play
the same role \cite{MMN} for ${\rm tr}\left(X \frac{\p}{\p X}\right)^n$}, i.e. to the ray $(0,1)$

\item{Further generalization to other rays $(m,n)$}

\item{The theory of unitary correlators and its $\beta$ and $(q,t)$ deformations}

\end{itemize}

\section*{Acknowledgements}

We appreciate numerous discussions on the subjects of this paper with Yaroslav Drachov, Dmitriy Galakhov, Fan Liu, 
Andrei Mironov,  Victor Mishnyakov, Alexander Popolitov, Shamil Shakirov, Nikita Tselousov, Rui Wang and Wei-Zhong Zhao.

This work is partly supported by the RSF Grant No.24-12-00178.

\end{document}